\documentclass[a4paper, 12pt]{article} 

\usepackage{amsfonts}
\usepackage{amssymb}
\usepackage{latexsym}

\title{Bohmian Trajectories Within Hilbert Space Based Quantum Mechanics. 
Solution of the Measurement Problem}

\author{Tulsi Dass}

\date{}

\begin{document}

\maketitle

de Broglie-Bohm theory (dBBT), treating quantum particles as point ob\-jects moving along well defined (Bohmian) trajectories, offers an appealing solution of the measurement problem in quantum mechanics; it has, however, problems relating to spin, relativity and lack of proper integration with the Hilbert space based framework. In this work, we present a consistent Hilbert space based formalism which has the traditional state-observable framework integrated with the desirable features of dBBT. We adopt ensemble interpretation for the Schr$\ddot{o}$dinger wave function $\psi$. Given a Schr$\ddot{o}$dinger wave function $\psi$, we use its value $\psi_{0}$ at some fixed time (say, t = 0) to define the probability measure $|\psi_{0}|^2 dx$ on the system configuration space $ M (= \mathbb{R}^n)$. On the resulting probability space $\mathcal{M}_{0}$, we introduce a stochastic process $\xi(t)$ corresponding to the Heisenberg position operator $X_{H}(t)$ such that, in the Heisenberg state $|\psi_{H}\rangle $ corresponding to $\psi_{0}$, the expectation value of $X_{H}(t)$ equals that of $\xi(t)$ in $\mathcal{M}_{0}$. This condition leads to the de Broglie-Bohm  guidance equation for the sample paths of the process $\xi(t)$ which are, therefore, Bohmian trajectories supposedly representing time-evolutions of individual members of the $\psi_{0}$-ensemble. Stochastic processes and Bohmian trajectories corresponding to observables with discrete eigenvalues (in particular spin) are treated by extending the configuration space to the spectral space of the commutative algebra obtained by adding appropriate discrete observables to the position observables. Pauli's equation (the Schr$\ddot{o}$dinger equation for a nonrelativistic charged spin half particle) is treated as an example. A straightforward \emph{derivation} of von Neumann's projection rule employing the Schr$\ddot{o}$dinger - Bohm evolution of individual systems along their Bohmian trajectories is given. Some comments on the potential application of the formalism developed here to quantum mechanics of the universe are included. 

\vspace{4 mm}

\vspace{12 mm}
\noindent \textbf{Contents}
\begin{description} 
\item[\textbf{1}] Introduction
\item[\textbf{2}] Background \\
\textbf\small{2.1} Some lessons from Hamilton-Jacobi theory \\
\textbf\small{2.2} Measurement problem in quantum mechanics \\ 
\textbf\small{2.3} de Broglie-Bohm theory
\item[\textbf{3}] Integration of pilot wave concept with the Hilbert space based framework of quantum mechanics \\
\textbf\small{3.1} Stochastic process corresponding to the Heisenberg position operator \\
\textbf\small{3.2} Relationship between Bohmian and Feynman trajectories \\
\textbf\small{3.3} Values of general observables along Bohmian trajectories \\
\textbf\small{3.4} Generalization to observables with discrete eigenvalues \\
\textbf\small{3.5} Symmetry related considerations 
\item[\textbf{4}]Bohmian trajectories for particle position and spin related to Pauli's equation 
\item[\textbf{5}] Derivation of von Neumann's projection rule based on S-B evolution of individual systems
\item[\textbf{6}] Concluding remarks
\end{description}

\vspace{4 mm} 
\section{Introduction} 

An awkward feature of the traditional formalism of quantum mechanics (QM) is the so called measurement problem. In the standard quantum mechanical treatment of measurement of an observable F of a system $\cal{S}$, interaction between the system $\cal{S}$ and apparatus $\cal{A}$ yields a superposition of product states of the eigenvectors of F with the corresponding pointer states. [See Eq (26) below.] Each term in the expansion corresponds to a possible result of the measurement. To ensure a unique result in each individual run of the experiment, an ad hoc reduction/collapse postulate [1] is invoked which serves to randomly select one of the terms in the superposition. The measurement problem mentioned above is essentially the problem of finding a rational explanation of the reduction in a proper theoretical framework.

An appealing explanation of the reduction is provided by the de Broglie-Bohm theory (dBBT) [2,3,4,5] which supliments, for the description of instantaneous state of a quantum system, the  Schr$\ddot{o}$dinger wave function with configuration space coordinates of the particles constituting the system; time evolution of the particle positions is governed by a guidance equation involving a velocity field determined by the wave function [see equations (30,35) below]; the resulting particle trajectories in configuration space are called Bohmian trajectories. In an individual run of an experiment, the Bohmian trajectory of the combined ($\cal{S}$ + $\cal{A}$) system serves to select (depending on initial position of the trajectory) an appropriate single term in the superposition; this happens because pointer states of $\cal{A}$ are macroscopically distinguishable. This theory, however, has its own problems (relating to the treatment of spin and relativity and to lack of proper integration of Bohmian trajectories related developments with the Hilbert space based formalism) which have prevented its acceptance in the mainstream of quantum physics.

We propose to integrate the tra\-ditional state-observable for\-malism of QM with desirable features of dBBT in a consistent Hilbert space based probabilistic framework. Adopting en\-semble interpretation for the Schr$\ddot{o}$dinger wave function $\psi$, we shall use $\psi$-stochastics in configuration space for the treatment of individual members of the $\psi$-ensemble; central objects in this treat\-ment will be stochastic processes constructed in this dy\=namical probability setting in correspondence with Heisenberg operators in the system Hilbert space. 

As in the main development of dBBT, we consider, in the present work,  only nonrelativistic quantum particle systems; the configuration space M of a typical quantum system will, therefore, be taken as $\mathbb{R}^{n}$. Given a wave function $\psi$(x,t), we use the fixed time function $\psi_{0}(x) = \psi (x,0)$ to construct a probability space (M, $|\psi_{0}|^{2}$dx). On this probability space, we introduce an M-valued stochastic process $\xi(t)$ corresponding to the Heisenberg position operator $X_H(t)$; its sample paths $\xi(t,x) = \xi(t)(x)$ satisfying the condition 
\begin{equation}
\xi(0,x) = x , \ \ x \in M 
\end{equation}
corresponding to different x are supposed to represent trajecto\-ries of individual sys\-tems in the $\psi$-ensemble. Stochastics of $\xi(t)$ is proposed to be related to that of $X_H(t)$ by demanding the equality of their expectation values in their respective frame\-works for all times t : 
\begin{equation}
\int_{M} \xi(t,x) |\psi_0(x)|^2 dx = \langle\psi_H | X_H (t) |\psi_H \rangle 
\end{equation}
where $|\psi_H\rangle$ is the Heisenberg state vector corresponding to the Schr$\ddot{o}$dinger wave function $\psi_0$. Smoothness of Schrodinger evolution of the wave function $\psi(x,t)$, combined with Eq(2), ensures that the process $\xi(t)$ is diffentiable; its smooth sample paths satisfying the condition (1) are the appropriate Bohmian trajectories. 

The kinematic variables position, momentum, angular momentum etc all have, at each instant, simultaneously definite values on a Bohmian trajectory; these variables are referred to as be-ables (generally written simply as beables)[6] in the dBBT literature. This name reflects the understanding that particles on Bohmian trajectories \emph{have} these properties and not just \emph{are found} to have them on measurements.

Equations like (2), which have quantum mechanical observables and state vectors on one side and be-ables and probability measures on the other, serve to \newline 
(i) integrate the de Broglie's pilot wave idea with the traditional Hilbert space based framework of quantum mechanics (QM) [in particular, it yields, with Schr$\ddot{o}$dinger equation, the guidance equation of dBBT ; see Eq(44) below],\newline 
(ii) provide a relation between Bohmian trajectories and Fenman trajectories in the path integral based formalism [7], and \newline 
(iii) provide clues for solutions of problems arising from shortcomings of dBBT (in particular, indicate as to which symmetries in the relevant Hilbert space based set up are expected to be preserved in the corresponding be-able dynamics). 

Spin problem of dBBT is solved by generalizing the configuration space (which is the spectral space of the commutative algebra generated by the position operators) to the spectral space of a larger commutative algebra obtained by adding appropriate spin observables to the position observables.  Treatment of Pauli's equation in section 4 below provides an example of a system in which the above mentioned commutative algebra contains observables with continuous eigenvalues as well as those with discrete eigenvalues.          

The usual argument of invoking appropriate Bohmian trajectories to ensure unique outcomes in individual runs of an experiment proceeds as in dBBT. In section 5, we present a straightforward derivation of the von Neumann projection rule for the (system + apparatus) density operator [see Eq(29) below] based on the Schr$\ddot{o}$dinger-Bohm evolution ( S-B evolution, for brevity) of individual systems in an ensemble. 

Plan of the remaining sections is given in the table of contents above. In section 2.1, some insights and construction of the stochastic process $\zeta(t)$ are contributions of the present author. Sections 3,4 and 5 cover in detail the developments described above. The last section includes a remark on the suitability of the formalism developed here for the description of quantum mechanics of the universe. 

\par
\noindent \textsl{Remark} The idea of obtaining Bohmian trajectories as sample paths of a stochastic process along the lines mentioned above was first presented by the author back in 2005 [8]. Eq(2) above is essentially the same as the first equality in Eq(17) in [8]. At that time, however, the author believed that Bohmian like trajectories should be continuous and non-differentiable and expected an additional Brownian motion type term on the right in the guidance equation (30) below. The argument given for such an equation presented there ( Eq(24) in [8]) was, unfortunately (or, rather fortunately !), faulty and the referee of a respected journal rightly insisted that I should provide a proper argument for the additional term. The author realised quite recently that, in the same framework, the Bohmian trajectories can be rigorously shown to be differentiable. (This is essentially the consequence of smoothness of Sghr$\ddot{o}$dinger evolution of the quantum mechanical wave function.) The deeper level system trajectories in configuation space which correspond to quantum mechanical evolution are Feynman trajectories [7] which are known to be generally continuous and non-differentiable. The relation between Bohmian And Fenman trajectories treated in section 3.2 below involves some averaging operations which are responsible for smoothness of the former.   

\vspace{3mm} 
\section{Background} 

\subsection{Some lessons from Hamilton-Jacobi theory} 

Although good reviews of Hamilton-Jacobi (H-J) theory relevant for quantum physics [9,5] are available, this subsection has been included to emphasize some aspects which will help bring out clearly certain points in quantum physics. 

A good starting point is the expression for general variation of the action $I = \int Ldt$ in classical mechanics [10] 
\begin{equation} 
\delta I = \int_{t_{1}}^{t_{2}} [L]_{\alpha} \overline{\delta} q^{\alpha} dt + [p_{\alpha}\delta q^{\alpha} - H \delta t]_{t_{1}}^{t_{2}} 
\end{equation} 
where symbols have the usual meaning and
\[ [L]_{\alpha} = \frac{\partial L}{ \partial q^{\alpha} } - \frac{D}{dt} (\frac{\partial L}{\partial \dot{q}^{\alpha}} )  \  \alpha = 1,..,n. \]
The variations are defined as
\begin{eqnarray*}
t \to t^{\prime} = t + \delta t , \ \ q^{\alpha}(t) \to q^{\alpha^{\prime}} (t^{\prime}) = q^{\alpha}(t) + \delta q^{\alpha}(t); \\ 
\bar{\delta}q^{\alpha}(t) = q^{\alpha^{\prime}}(t) - q^{\alpha}(t) = \delta q^{\alpha}(t) - \dot{q}^{\alpha} \delta t. 
\end{eqnarray*}
\noindent While applying Hamilton's principle, one takes $\delta t = 0$ and $ \delta q^{\alpha} (t) = 0 $ at $ t = t_{1} , t_{2} $ . The stationary action condition $ \delta I = 0 $ then gives the Lagrange's equations of motion $ [L]_{\alpha} = 0. $ For $ q^{\alpha}$ satisfying the equations of motion, the integral in Eq(3) vanishes; form of the boundary term indicates that the action, evaluated for such $ q^{\alpha}$, is a function of $ q^{\alpha}(t_1), q^{\alpha} (t_2), t_1 $ and $ t_2 $; call this function $S(q(t_{2}), t_{2} ; q(t_{1}), t_{1})$. Keeping $t_{1}$ and $ q(t_{1})$ fixed, replacing $t_{2}$ and $q(t_{2})$ by t and q and denoting the resulting function as S(q,t), we have 
\[ \frac{\partial S(q,t)}{\partial q^{\alpha}} = p_{\alpha}(t); \ \ \frac{\partial S(q,t)}{\partial t} = - H(q,p,t) \] 
which leads to the H-J equation 
\begin{equation} 
\frac{\partial S(q,t)}{\partial t} + H \left( q, \frac{\partial S(q,t)}{\partial q}, t \right) = 0. 
\end{equation} 
A solution S(q,t) of Eq(4) with the initial condi\-tion $ S(q, t_{0}) = S_{0}(q)$, when supplemented with the initial condition 
\begin{equation} 
q^{\alpha} (t_{0}) = q^{\alpha}_{0} 
\end{equation}
leads to the unique trajectory ( q(t), p(t)) in phase space corresponding to the initial con\-dition $(q(t_{0}), p(t_{0})) = (q_{0}, p_{0})$ where 
\[ p_{\alpha 0} =\left. \frac{\partial S_{0}(q)}{\partial q^{\alpha}} \right|_{ q = q_{0}} \] 
as follows: Define momentum and velocity fields p(q,t) and v(q,t) by 
\begin{equation} 
p_{\alpha}(q,t) = \frac{\partial S(q,t)} { \partial q^{\alpha}}, \  v^{\alpha} (q,t) = \left. \frac{ \partial H(q,p,t)}{\partial p_{\alpha}} \right|_{p = p(q,t)}. 
\end{equation} 
We obtain q(t) from the differential equation 
\begin{equation} 
\frac{dq^{\alpha}}{dt} = v^{\alpha} (q(t),t) 
\end{equation} 
with the initial condition (5) and $ p_{\alpha}(t) = p_{\alpha}(q(t),t). $ 

The function S(q,t) is the state function of a dynamical system whose time evolution is governed by the H-J equation (4). Its physical significance for the system described by the action I is that, at any fixed t, it contains kinematic information about the dynamical trajectories passing through a general point q in the configuration space at time t. [See equations (6,7).] 

Formally, this function may be considered as the state function of an ensemble of systems, where different members of the ensemble correspond to different values of $q_{0}$ in Eq (5). To obtain the trajectory of an individual system in this ensemble, one must specify the solution S(q,t) of the H-J equation (4) [corresponding to an appropriate initial value $S_{0}(q)$] as well as the initial value $q_{0}$ in Eq(5). It follows that, at a fixed time $t_{0}$, whereas the state of the ensemble is represented completely  by the function $S_{0}$, that of an individual system in the ensemble is represented by the pair $ (S_{0}, q_{0})$. This observation will be helpful in understanding the status of the Schr$\ddot{o}$dinger wave function $\psi$(q,t) in a later section. 

Note the contrast between the types of description of classical dynamics in terms of an action functional I[q] on the one hand and the H-J function S(q,t) on the other. The generalized coordinates, velocities and momenta appearing in the first description describe the kinematics and dynamics of single concrete systems; they are be-ables as defined above. The H-J function S(q,t) describes, as we have seen above, an ensemble of systems. The derivation of the H-J equation starting from the action functional presented above is an example of starting with a description of dynamics of a system in terms of relevant be-ables and obtaining one in terms of H-J type functions which are time dependent fields defined on the system configuration space. In dBBT (section 2.3) and in section 3 below, one is doing the opposite. 

If, instead of the condition (5), we are initially given a probability distribution $\rho(q, t_{0}) = \rho_{0}(q)$ on the configuration space, we shall obtain, instead of a unique trajectory, a probability density function $\rho(q,t)$ satisfying the probability conservation equation 
\begin{equation} 
\frac{\partial \rho(q,t)}{\partial t} + \frac{\partial}{\partial q^{\alpha}}[ v^{\alpha} (q,t) \rho (q,t)] = 0. 
\end{equation} 
This can be obtained by applying the Liouville equation for the phase space probability density $\rho_{ph}$(q,p,t)
\[ \frac{\partial \rho_{ph}}{\partial t} + \frac{\partial \rho_{ph}}{\partial q^{\alpha}} \frac{\partial H}{\partial p_{\alpha}} - \frac{\partial \rho_{ph}}{\partial p_{\alpha}} \frac{\partial H}{\partial q^{\alpha}} = 0 \] 
to the function 
\[ \rho_{ph} (q,p,t) = \rho(q,t) \prod_{\alpha =1}^{n} \delta \left( p_{\alpha} - \frac{\partial S(q,t)} { \partial q^{\alpha}} \right)  \] 
and making use of equations (4,6). The dynamical system with state functions S(q,t) and $\rho(q,t)$ satisfying the evolution equation equations (4,8) is often referred to as `Hamilton-Jacobi fluid'. 

In this probabilistic extension of H-J theory, the concept of configuration space trajectories remains meaningful; we only do not have \emph{complete} information about them. Probabilistic questions relating to (families of) such trajectories can be answered in the formalism. In the limiting case of the initial probability density $\rho_{0}(q) = \delta(q,q_{0})$, we recover the trajectories for individual systems. If we insist on $\rho$ being a well defined function, such trajectories can be realized arbitrarily closely but not exactly. 

At this point, it is instructive (in anticipation of some similar developments later) to exploit the differentiability of the density function $\rho$ in H-J fluid to introduce a stochastic process which has smooth sample paths satisfying the trajectory equation (7). To this end, considering the configuration space M(= $\mathbb{R}^n$) (with its Borel $\sigma$-algebra) as a probability space with probability measure $\rho_0(q)$dq, we introduce an M-valued  stochastic process $\zeta$(t) such that its probability density function is $\rho$(q,t). The expectation value of $\zeta$(t) is
\begin{equation} 
E[\zeta^{\alpha}(t)] = \int_M \zeta^{\alpha}(t,q^{\prime})\rho_{0}(q^{\prime})dq^{\prime} = \int_M q^{\alpha} \rho(q,t)dq. 
\end{equation}                
Now, using Eq(8),    
\begin{eqnarray*} 
E[\zeta^{\alpha}(t +dt) - \zeta^{\alpha}(t)] & = & dt \int_M q^{\alpha}\frac{\partial \rho(q,t)}{\partial t}dq + O(dt^{2}) \\
& = & dt \int_M v^{\alpha}(q,t)\rho(q,t)dq + O(dt^{2}) 
\end{eqnarray*} 
where, after an integration by parts, the surface term has been dropped. This gives 
\begin{equation}
E[\zeta^{\alpha}(t+dt) - \zeta^{\alpha}(t) - v^{\alpha}(\zeta(t), t)dt] = O(dt^{2}) 
\end{equation}
which implies that the stochastic process $\zeta(t)$ is differentiable in the sense of convergence in measure [11] and satisfies the stochastic differential equation (SDE)
\begin{equation} 
\frac{d\zeta^{\alpha}(t)}{dt} = v^{\alpha}(\zeta(t),t), \ \  \alpha = 1,..., n.
\end{equation} 
Its sample paths $ \zeta (t)(q) = \zeta(t,q)$ satisfy the differential equation 
\begin{equation} 
\frac{\partial \zeta^{\alpha}(t,q)} {\partial t} = v^{\alpha}(\zeta(t,q), t). 
\end{equation}
Imposing the initial condition 
\begin{equation}
\zeta(t_{0}, q) = q, 
\end{equation} 
we see that the functions 
\begin{equation}
q^{\alpha}(t) = \zeta^{\alpha}(t,q_{0}) 
\end{equation}
satisfy equations (7,5) and, therefore represent the dynamical trajectories of the system under consideration.

In section 3, we shall construct a similar stochastic process $\xi$(t) associated with the Heisenberg position operator $X_H$(t) of a quantum particle; its sample paths satisfying a condition analogous to (13) will be appropriate Bohmian trajectories. The role played by the evolution equations of the H-J fluid [equations (4,8)] above will be played by the Schr$\ddot{o}$dinger equation which is closely related to the former, as is clear from the example below.  

For a nonrelativistic spinless particle with the Hamiltonian 
\begin{equation} 
H(\mathbf{x,p}) = (2m)^{-1} |\mathbf{p}|^{2} + V(\mathbf{x}) 
\end{equation}
the Hamilton's equation are
\[ \frac{d\mathbf{x}}{dt} = m^{-1}\mathbf{p}(t), \ \   \frac{d\mathbf{p}(t)}{dt} = - (\mathbf{\nabla}V)(\mathbf{x}(t)) . \]    
Equations for the corresponding H-J fluid described by the functions S(\textbf{x},t) and $\rho$(\textbf{x},t) are 
\begin{eqnarray}
\frac{\partial S(\mathbf{x},t)}{\partial t} + (2m)^{-1} |(\mathbf{\nabla}S)(\mathbf{x},t)|^{2} + V(\mathbf{x}) = 0 \\
\frac{\partial \rho(\mathbf{x},t)}{\partial t} + \mathbf{\nabla}.[(\rho \mathbf{v})(\mathbf{x},t)] = 0 
\end{eqnarray}
where
\begin{equation}
\mathbf{v}(\mathbf{x},t) = m^{-1} (\mathbf{\nabla}S)(\mathbf{x},t). 
\end{equation} 
Given a solution S(\textbf{x},t) of Eq(16),the individual system trajectories are sample paths of the stochastic process $\mathbf{\zeta}$(t) governed by the SDE
\begin{equation}
d\zeta_{j}(t) = \frac{1}{m} \left. \left( \frac{\partial S (\mathbf{x},t)}{\partial x_{j}} \right) \right|_{\mathbf{x} = \mathbf{\zeta}(t)} dt  \ \  j = 1,2,3.
\end{equation} 

The probabilistc extension of H-J theory presented above has a close relation to the wave mechanics based on the Schr$\ddot{o}$dinger equation
\begin{equation} 
i\hbar \frac{\partial \psi (\mathbf{x},t)} {\partial t} = [ - \frac{\hbar^{2}}{2m}\nabla^{2} + V(\mathbf{x})] \psi(\mathbf{x},t) . 
\end{equation} 
Putting $ \psi(\mathbf{x},t) = \tilde{R}(\mathbf{x},t) exp[i \tilde{S}(\mathbf{x},t)/\hbar]$ in Eq(20), the real and imaginary parts of the resulting equation give (with $ \tilde{\rho} = \tilde{R}^{2}$) 
\begin{eqnarray} 
\frac{\partial \tilde{\rho}}{ \partial t} + \mathbf{\nabla}.(\tilde{\rho}\tilde{\mathbf{v}}) = 0 \ \ \mbox{where} \ \ \tilde{\mathbf{v}} = m^{-1}\mathbf{\nabla} \tilde{S} \\
\frac{\partial \tilde{S}}{\partial t} + \frac{|\mathbf{\nabla} \tilde{S}|^{2}}{2m} + V - \frac{\hbar^{2}}{2m} \frac{\nabla^{2} R}{R}  = 0. 
\end{eqnarray} 
The functions $\tilde{\rho}$ and $\tilde{S}$ may have some $\hbar$-dependence. Assuming they have smooth $\hbar \to 0$ limits (respectively $\rho$ and S), equations (21,22) give, in the $\hbar \to 0$ limit, equations (17) and (16).

An important point worth noting here is that, whereas in the augmented H-J theory two real valued functions S and $\rho$ have separate roles of describing, respectively, dynamics  and stochastics of particle motion, in Schr$\ddot{o}$dinger theory we employ a single complex valued function $\psi$ to describe both dynamics and stochastics.

Such a unification of probability and dynamics [considered analogous to the unification of electric and magnetic fields in electromagnetic field tensor] was proposed in [12]  as the fundamental unifying principle underlying QM. It was shown there that, when such a unification is attempted at the configuration space level, the appropriate single mathematical object unifying the H-J function S(q,t) and the probability density $\rho$(q,t) must be a Schr$\ddot{o}$dinger type wave function which is `essentially'
\ensuremath{\sqrt{\rho}exp[i\lambda S]} with $\lambda$ real. This provides (through an admittedly heuristic reasoning) a logically satisfying way of introducing the wave function $\psi$ and the Planck constant ($\hbar = \lambda^{-1}$); the probability interpretation of $\psi$ is automatic.

\par
\noindent \textsl{Remark} To have a clear idea of the setting in which the unification is attempted and see that the argument involved is not quite trivial, the reader is advised to have a look at sections II and III of [12].     

\subsection{Measurement problem in quantum mechanics} [1,13,14,15] 

Before we state the problem, we present below a brief account of measurements in QM following essentially [1].

We shall treat both-the sytem of interest $\mathcal{S}$ and apparatus $\mathcal{A}$ as quantum systems with respective Hilbert spaces $\mathcal{H}_{\mathcal{S}}$ and $\mathcal{H}_{\mathcal{A}}$; the joint Hilbert space is the tensor product $\mathcal{H} = \mathcal{H}_{\mathcal{S}}\otimes \mathcal{H}_{\mathcal{A}}$. We seek measurement of an observable of $\mathcal{S}$ represented by a self adjoint operator F which, for simplicity, is assumed to have a discrete nondegenerate spectrum, the corresponding eigenvalue equation being $F|\psi_{j}\rangle = \lambda_{j} |\psi_{j}\rangle.$ Initial state of the system is asumed to be given as a finite superposition
\begin{equation}
|\psi_{in}\rangle  =  \sum_{j} c_{j} |\psi_{j}\rangle 
\end{equation} 
where $c_{j}$ are complex numbers. The apparatus $\mathcal{A}$ is supposedly equipped with macroscopically distinguishable pointer positions in one-to-one correspondence with the eigenvalues of F. These are represented by states $|\phi_{j}\rangle$ in $\mathcal{H}_{\mathcal{A}}$; the corresponding wave functions $\phi_{j}(y) = \langle y|\phi_{j}\rangle$ have, for different j, substantially non-overlapping supports in the apparatus configuration space coordinated by y.

We consider an ideal measurement, i.e. one in which the system of interest is not destroyed and which, on repetition, gives the same result. Denoting the `ready' state of the apparatus by $|\phi_{0}>$, the unitary operator U representing the effect of the $\mathcal{S}-\mathcal{A}$ interaction acts as 
\begin{equation}
U ( |\psi_{j}\rangle |\phi_{0}\rangle) = |\psi_{j}\rangle |\phi_{j}\rangle. 
\end{equation}
Linearity of the operator U implies that, for the initial system-apparatus state
\begin{equation}
|\Psi_{in}\rangle = \sum_{j} c_{j} |\psi_{j}\rangle |\phi_{0}\rangle, 
\end{equation}
we have the final state 
\begin{eqnarray}
|\Psi_{f}\rangle & = & U|\Psi_{in}\rangle  \nonumber \\ 
           & = & \sum_{j} c_{j} |\psi_{j}\rangle|\phi_{j}\rangle;
\end{eqnarray}
the corresponding wave function is [with $\psi_{j}(x) = \langle x|\psi_{j}\rangle$ where x labels points in the system configuration space]
\begin{eqnarray} 
\Psi_{f}(x,y) & = & \langle x,y|\Psi_{f}\rangle    \nonumber  \\ 
              & = & \sum_{j} c_{j} \psi_{j}(x) \phi_{j}(y). 
\end{eqnarray}
  
 A non-trivial superposition of the type (27) cannot be an acceptable final result of a measurement because, in each run of an experiment, one always obtains a definite pointer reading (which may be  different in different runs of the experiment). To explain this, von Neumann [1] introduced an ad hoc postulate (commonly called reduction/collapse postulate) which states that, at this stage, a non-causal and discontinuous change takes place (supplementing the Schr$\ddot{o}$dinger evolution U) whose effect is to randomely select one term in the superposition (26) and delete the others; this amounts to replacing the density operator 
 \begin{equation}
 \rho^{(\Psi_{f})} = |\Psi_{f}\rangle\langle\Psi_{f}| = \sum_{j,k} c_{j}c_{k}^{*} (|\psi_{j}\rangle\langle\psi_{k}|)(|\phi_{j}\rangle\langle\phi_{k}|)
 \end{equation}
 by the reduced density operator
 \begin{equation} 
 \rho_{red} = \sum_{k} P_{k} \rho^{(\Psi_{f})} P_{k} = \sum_{j} |c_{j}|^{2}(|\psi_{j}\rangle\langle\psi_{j}|)(|\phi_{j}\rangle\langle\phi_{j}|) 
 \end{equation} 
 where $ P_{k} = |\psi_{k}\rangle\langle\psi_{k}|$. In multiple runs, probability of the state $|\psi_{j}\rangle|\phi_{j}\rangle$ appearing as the final apparatus-system state is $p_{j} = |c_{j}|^{2}$.
 
 Predictions made on the basis of Eq(29) are always in agreement with experiment. The `measurement problem in QM' is essentially the problem of replacing the ad hoc reduction/collapse postulate by a rational explanation of the reduction above.
 
 \subsection{de Broglie-Bohm theory} [2,3,4,5,16] 
 
 The de Broglie-Bohm the\-ory(dBBT) (also called Bohmian mechanics) treats particles participating in microscopic 
 level processes as concrete physical point-like en\-tities moving along well defined trajectories. A nonrelativistic 
 particle has associ\-ated with it a com\-plex \emph{physical} field $\psi$(\textbf{x},t); this field and the 
 trajectory function \textbf{r}(t) define completely the state of the particle-field system at time t. Evolution 
 equations for the state are the Schr$\ddot{o}$dinger equation(20) and the deBroglie-Bohm guidance equation
 \begin{equation}
 \frac{d\mathbf{r}(t)}{dt} = \mathbf{v}^{(\psi)}(\mathbf{r}(t),t) 
 \end{equation} 
 where the velocity field $\mathbf{v}^{(\psi)}$(\textbf{x},t) is 
 \begin{eqnarray}
 \mathbf{v}^{(\psi)}(\mathbf{x},t) & = & \left(\frac{\mathbf{J}^{(\psi)}}{\psi^{*}\psi}\right)(\mathbf{x},t) \nonumber \\
 & \equiv & \frac{\hbar}{m}\left[\frac{\Im {(\psi^{*}\mathbf{\nabla}\psi)}}{\psi^{*}\psi}\right](\mathbf{x},t). 
 \end{eqnarray} 
 Here $\mathbf{J}^{(\psi)}$ is the `probability current' appearing in the probability conservation equation 
 \begin{equation}
 \frac{\partial |\psi(\mathbf{x},t)|^{2}}{\partial t} + (\mathbf{\nabla}.\mathbf{J}^{(\psi)})(\mathbf{x},t) = 0. 
 \end{equation} 
 
 Equations (20) and(30) constitute a deterministic system of evolution equations admitting, with initial conditions 
 \[ \psi(\mathbf{x},0) = \psi_0(\mathbf{x}), \ \ \mathbf{r}(0) =\mathbf{x}_0, \] 
 a unique solution in appropriate domain. Through equations (30,31) the wave functions $\psi$ serves as a `guidance field' for the particle (giving concrete shape to de Broglie's `pilot wave' idea). Putting $\psi = R exp[iS/\hbar]$ in Eq(20), we obtain two real variable equations (21,22) where the last term in Eq(22) is the so called `quantum potential' 
 \begin{equation}
 V_Q = - \frac{\hbar^{2}}{2m}\frac{\nabla^{2} R}{R}.
 \end{equation} 
 The quantum potential is employed in dBBT to give realistic descriptions of atomic processes involving Bohmian trajectories of atomic electrons. 
 
  For N nonrelativistic spinless particles, with the wave function $\psi(\mathbf{x}_{1},..,\mathbf{x}_{N};t)$ and trajectory functions $\mathbf{r}_{a}(t)$(a = 1,..,N), analogues of equations (20,30,31) are
 \begin{eqnarray}
 i \hbar \frac{\partial \psi}{\partial t} = \left[- \sum_{a=1}^{N} \frac{\hbar^{2}}{2m_{a}}\nabla_{a}^{2} + V(\mathbf{x}_{1},..,\mathbf{x}_{N})\right] \psi \\
 \frac{d\mathbf{r}_{a}(t)}{dt} = {{v}_{a}}^{(\psi)}(\mathbf{x}_{1},..,\mathbf{x}_{N};t) \rfloor_{\mathbf{x}_{a} = \mathbf{r}_{a}(t)} \ \ a = 1,..,N \\ 
 \mathbf{v}_{a}^{(\psi)}(\mathbf{x}_{1},..,\mathbf{x}_{N};t) = \frac{\hbar}{m_{a}}\Im \left(\frac{\psi^{*}\mathbf{\nabla}_{a}\psi}{\psi^{*}\psi} \right) \ \ a = 1,..,N 
 \end{eqnarray}
 Note that, in Eq(35), the velocity function of any particle depends, at any time t,  generally on the instantaneous positions of all the N particles --- a manifestation of nonlocality of QM.
 
 Appearance of probability in dBBT differs at a conceptual level from the traditional formalism of QM in two respects : \newline 
 (i) The quantity $|\psi(\mathbf{x})|^{2}\Delta V $ is understood in dBBT to be the probability of the particle \emph{being} in the volume element $\Delta V$ whereas, in the traditional formalism, it is understood to be the probability of the same \emph {being found} in $\Delta V$ on a position measurement. \newline 
 (ii) In the traditional framework, probability is understood to be intrinsic for the theory; dynamics of a quantum system is irreducibly probabilistic. In dBBT, however, the guidance equa\-tion (30) [along with the Schr$\ddot{o}$dinger equation (20)] permits the particle, given an initial position $\mathbf{x}_{0}$, to have a well defined trajectory. Need for introducing probability arises because, at the level of observations, precise initial configuration of a microscopic system is not under control. 
 
 dBBT provides a solution to the measurement problem by taking into consideration, in the treatment of measurement on a quantum system, the Bohmian trajectories of the system and the apparatus. In a single run of the experiment, corresponding to their initial configuration values, one has a unique Bohmian trajectory in the combined (system + apparatus) configuration space. Recalling Eq(27), since the wave packets $\phi_{j}$ corresponding to different values of j are practically non-overlapping, this trajectory will enter the support domain of only one of the joint packets $|\psi_{j}\rangle |\phi_{j}\rangle$ leading to a unique outcome in the experiment. Randomness of outcomes arises due to different possible initial values ($x_0,y_0$) of (system + apparatus) configurations (which are not under experimental control) in different runs of the experiment.
 
 Main unsatisfactory features of dBBT are as follows. 
 
 \par
 \noindent (i) It does not provide satisfactory treatment of be-ables corresponding to spin observables of particles [analogous to trajectory functions \textbf{r}(t) corresponding to the Heisenberg position observables $\mathbf{X}$(t)]. The two relevant chapters in the book [5] contain much instructive material; however, the only presently relevant outcome (in [5] section 9.3) is the detailed treatment of the Bohmian electron trajectory guided by a solution of Pauli's equation [Eq(50) below] in which effect of spin on the motion of an individual electron is reflected through the properties of the spinor wave function. The book ([16] page 165) has this to say about spin : 
 \begin{quote}
 ``Spin is a property of the guiding wave function which is a spinor. The particle itself has only one property : position.'' 
 \end{quote}
This is not acceptable. According to Eq(24), when $\psi$ is an eigenvector of an operator F, the particle/system \emph{has} the corresponding eigenvalue as the value of the observable represented by F. The squared spin operator $|\mathbf{S}|^{2}$ has a fixed eigenvalue $s(s+1)\hbar^{2}$ (s = 1/2 for electron, proton,..) for every spinor in the relevant irreducible representation of the spin algebra. This means that the particle \emph{has} this value for the observable corresponding to the operator $|\mathbf{S}|^{2}$ for all admissible wave functions; this value must, therefore, be taken as characteristic of the particle. In particular, an electron \emph{is} a spin 1/2 particle. 

\par
\noindent (ii) In the relativistic QM of particle systems, the augmented formalism including Bohmian tra\-jectories is Lorentz invariant only in the case of single particle Dirac equation (admitting both positive and negative energy solutions of the equation [5]). In the multiparticle case, nonlocal correlations between motions of all the particles violate Lorentz invariance; however, Lorentz invariance is restored in all statistical predictions. Efforts to construct a Lorentz invariant version of dBBT have not been successful.

For understanding the effective symmetries in a theory involving both observables and be-ables, one needs to examine the effect of symmetry operations in a proper setting involving state vectors, operators and Bohmian trajectory functions. This setting is provided by Eq(39) below. Further discussion on this point is postponed to the next section. 

\par
\noindent (iii) In dBBT, the wave function $\psi$ is considered as a \emph{physical} field analogous to the electromagnetic field in classical electrodynamics. This is not acceptable due to the following two reasons : 
\begin{enumerate}
\item The $\psi$ field affects the particles, but is not affected by them.
\item If $\psi$ is a physical field, then any limiting form of it (in paricular, the classical limit) should also be a physical field. However, the quantities R and S appearing in the polar decomposition of $\psi$ considered earlier become, in the classical limit, the quantities $\sqrt{\rho}$ and the H-J function (in the probabilistic extension of H-J theory) which are certainly not physical fields. 
\end{enumerate} 

\noindent (iv) Lack of proper integration of the Bohmian trajectories and related developments with the traditional Hilbert space based formalism. Bohm himself [3] referred to the trajectory functions as hidden variables. In later works, there is generally an overemphasis on the role of be-ables and traditional quantum mechanical observables are treated as secondary objects.(See, for example, [17,16].) In the book [16], title of section 12.1 is 
\begin{quote}
 Observables. An Unhelpful Notion 
 \end{quote}
As will be clear to the reader after going through sections 3 and 4 below, the problems of dBBT relating to spin are mainly due to not taking the algebra of observables seriously.   

\section{Integration of pilot wave con\-cept with the Hilbert space based framework of quantum mechanics} 

\subsection{Stochastic process corresponding to the Heisenberg position operator} 

We first consider a nonrelativistic particle with wave function $\psi$(x,t) having the traditional probability interpretation (in which $\psi$ is understood to represent an ensemble). We envisage different individual systems in the $\psi$-ensemble as point particles moving along well defined trajectories in $\mathbb{R}^3$. Following steps similar to those related to the stochastic process $\zeta(t)$ in section 2.1, we shall construct a stochastic process (related to the Heisenberg position operator $X_{H}$(t) for the particle [8]) having these trajectories as its sample paths. 

Given a solution $\psi$(x,t) of the Schr$\ddot{o}$dinger equation (20), we have the probability space 
\begin{equation} 
\mathcal{M}_{0} = \left( \mathbb{R}^{3}, Borel(\mathbb{R}^{3}), |\psi_{0}(x)|^{2}dx \right) 
\end{equation} 
where $\psi_{0}(x) = \psi(x,0)$. On this probability space, we consider an $\mathbb{R}^{3}$-valued stochastic process $\xi(t)$ whose sample path $\xi_{x}(t) = \xi(t)(x) = \xi(t,x)$ satisfying the initial condition [recall Eq(13)] 
\begin{equation}
\xi(0,x) = x 
\end{equation} 
is supposed to represent the trajectory function  of the individual system in the $\psi$-ensemble with the starting point $\xi_{x}$(0) = x. Different members of the $\psi$-ensemble correspond to different starting points in the configuration space $\mathbb{R}^{3}$. 

The process $\xi$(t) supposedly corresponds to the Heisenberg position operator $X_{H}$(t) for the particle. To display this connection, we invoke/use consistency between the probabilistic treatments of particle dynamics at the ensemble and individual system levels; we demand that, at each time t, the mean of the random variable $\xi_{t}(x)$ be equal to the expectation of the operator $X_{H}$(t) in the Heisenberg state $\psi_{H}$ corresponding to the Schr$\ddot{o}$dinger wave function $\psi_{S}(0) = \psi_{0}$ : 
\begin{eqnarray}
E[\xi^{j}(t)] & = & \int_{\mathbb{R}^{3}} \xi^{j}(t,y) |\psi_{0}(y)|^{2}dy \nonumber \\
              & = & <\psi_{H}|X_{H}^{j}(t)|\psi_{H}>, \ j = 1,2,3. 
\end{eqnarray} 
Going to the Schr$\ddot{o}$dinger picture with $\langle y|\psi_{S}(t)\rangle = \psi(y,t)$, the right hand side of Eq(39) equals 
\begin{equation}
\langle\psi_{S}(t)|X_{S}^{j}|\psi_{S}(t)\rangle = \int_{\mathbb{R}^3} x^{j}|\psi(x,t)|^{2}dx 
\end{equation} 
showing that the probability density function for the process $\xi(t)$ is 
\begin{equation}
\rho_{\xi}(x,t) = |\psi(x,t)|^{2}. 
\end{equation} 
Now, recalling equations (20,32) 
\begin{eqnarray} 
E[\xi^{j}(t+dt) - \xi^{j}(t)] & = & \int x^{j} [ |\psi(x, t+dt)|^{2} - |\psi(x,t)|^{2}]dx  \nonumber \\ 
  & = & - dt \int x^{j} \mathbf{\nabla}. \mathbf{J}^{(\psi)}(x,t) dx + O(dt^{2})  \nonumber \\ 
  & = & dt \int J^{(\psi)j}(x,t)dx + O(dt^{2}). 
\end{eqnarray} 
Defining the velocity field $v^{(\psi)j}$ as in Eq(31), Eq(42) gives 
\begin{equation}
E[\xi^{j}(t+dt) - \xi^{j}(t) - v^{(\psi)j}(\xi(t),t)dt] = O(dt^{2}) 
\end{equation} 
showing that the process $\xi(t)$ is differentiable in the sense of convergence in measure, implying 
\begin{equation}
\frac{d\xi(t)}{dt} = v^{(\psi)}(\xi(t),t); 
\end{equation} 
the sample path $\xi_x(t)$ satisfies the Bohmian trajectory equation (30). 

Genralization of the develop\-ment above to the N-particle case is straightforward. Since 3-dimensionality of the configuration space has not been used in the steps above, we need only replace $\mathbb{R}^{3}$ in Eq(37) by $\mathbb{R}^{3N}$ and have j in Eq(39) take values 1,..3N [undestood as j= ka; k=1,2,3 and a = 1,.. N].

We shall refer to the evolution of individual systems in an ensemble governed by equations like (30) and (44) as \emph{Schr$\ddot{o}$dinger-Bohm evolution}. In fact, it is just the `extension' of the evolution of $\psi(t)$ at the ensemble level to the individual system state $(\psi(t), \mathbf{r}(t))$. If the $\psi(t)$ evolution is considered as a river flow, then the $(\psi(t), \mathbf{r}(t))$ evolution is analogue of the flow of a small floating object in the river. Recalling that, in dBBT, the guidance equation (30) was put `by hand' as evolution equation of a `hidden variable', its integration with the Hilbert space based formalism presented above serves to define its proper status in the formalism of quantum mechanics; in the process, it serves to bring out the beauty and depth/richness of Schr$\ddot{o}$dinger evolution. 

The trajectories (for the system and the apparatus) of such evolutions can now be used in the treatment of measurements on quantum systems as in section 2.3, thus providing a solution of the measurement problem without going outside the Hilbert space based framework. Having obtained the effective collapse of the (system + apparatus) state vector in individual runs of an experiment, one can obtain the result $p_{j} = |c_{j}|^{2}$ by appealing to the probability interpretation of the wave function $\Psi(x,y)$ of Eq(27)and using the orthogonality properties of the summands in it (as, for example, in [5] page 345) and obtain $\rho_{red}$ of Eq(29). A direct derivation of the von Neumann projection equation (29) by considering the combined effect of Schr$\ddot{o}$dinger-Bohm evolutions of all individual systems in the ensemble associated with the initial (system + apparatus) wave function on the initial (system + apparatus) density operator $\rho^{(\Psi_{in})} = |\Psi_{in}\rangle\langle\Psi_{in}|$ [see Eq(25)] will be given in section 5. 

\subsection{Relation between Bohmian and Fenman trajectories} 

Eq(39), which relates observables and state vectors with the cor\-responding be-ables and probability mea\-sures, serves, through the integral form of the evolution of Schr$\ddot{o}$dinger wave function employing path integrals, to provide relation between Bohmian and Feynman trajectories. We have 
\begin{eqnarray}
\psi(x,t) = \langle x|\psi_{S}(t)\rangle & = & \langle x|U(t,0)|\psi_{S}(0)\rangle  \nonumber \\ 
& = & \int\langle x|U(t,0)|y \rangle dy \langle y|\psi_{S}(0)\rangle  \nonumber \\ 
& = & \int K(x,t;y,0)\psi_{0}(y)dy; 
\end{eqnarray} 
\begin{equation} 
K(x,t;y,0) = \int_{paths (y,0) \to (x,t)} D\gamma exp[\frac{i}{\hbar} \int_{\gamma}Ldt]. 
\end{equation} 
Equations (39,40,45,46) provide the desired connection (admittedly not in a simple form). Note from these equations the important fact that, whereas Feynman trajectories are determined by the unitary evolution (through the Lagrangian or the Hamiltonian) only, Bohmian trajectories are determined jointly by the unitary evolution and the initial wave function. 

\subsection{Values of general observables along Bohmian trajectories} 

Given a self adjoint operator A in the Hilbert space of a quantum system, we have 
\begin{equation} 
\langle\psi_{S}(t)|A|\psi_{S}(t)\rangle = \int A(x,t) |\psi(x,t)|^{2}dx 
\end{equation} 
where 
\begin{equation} 
A(x,t) = \frac{1}{2}[\psi^{*}(A\psi)+(A\psi)^{*}\psi](x,t)/|\psi(x,t)|^{2} 
\end{equation} 
is the quantity called `local expectation value' of the observable A in [5]. Recalling equations (40,41), right hand side of Eq(47) is the expectation value of the random variable $A(\xi(t),t)$. It follows that the value of the observable A along the Bohmian trajectory $\xi_{x}(t)$ is $A(\xi_{x}(t),t)$ --- a result obtained in [5] by intuitive reasoning. Explicit expressions for these values for some observables using Eq(48) can be found in [5]. The ones more often used are, recalling the polar decomposition of $\psi(x,t)$ and Eq(22), those (for a spinless nonrelativistic particle) for the momentum ($P_{j} = \frac{\partial S}{\partial x^{j}}$) and energy (E = - $\frac{\partial S}{\partial t}$). 

\subsection{Generalization to observables with discrete eigenvalues}

To explore the possibility of a generlisation of the developments above to include spin type observables, it is useful to note that, in Eq(39), the operators \{$X_{j}$; j= 1,2,3\} generate a commutative algebra (call it the `configuration algebra') and have the Schr$\ddot{o}$dinger wave function $\psi$ as a common generalized eigenfunction; their spectral space $\mathbb{R}^{3}$ appears as as the probability space in Eq(37). This suggests that, for the desired generalization, one must replace the configuration algebra by its minimal abelian extension including the needful discrete observables. 

For a nonrelativistic particle with nonzero spin, the relevant complete set of commuting observables is (with traditional notation for the spin operators)  
\begin{equation} 
\mathcal{C} = \left( \{ X_{j}; j = 1,2,3 \}, |\mathbf{S}|^{2}, S_{3} \right). 
\end{equation}
Of these, $|\mathbf{S}|^{2}$, with eigenvalues s(s+1)$\hbar^{2}$ for all states for the particle, serves to label it with spin s. The desired abelian algebra in this case is the one generated by $\{X_{1}, X_{2}, X_{3}, S_{3} \}$. Treatment of Bohmian trajectories for this algebra will be given, for s = 1/2, in the context of Pauli equation in the next section. Generalization to higher s and more general discrete observables will be obvious. 

\par
\noindent \textsl{Note} Another complete set of commuting observables, mathemati\-cally equivalent to (49), is the one with \{ $X_{j}$ \} replaced by the momentum observables \{ $P_{j}$ \}. However, the set (49) is more important due to its relation with physical space and (consequently) its relevance for the consideration of existence/reality matters. 

\subsection{Symmetry related considerations} 

Recalling the problem with Lorentz invariance in dBBT, we analyse it here in the broader context of implementation of a symmetry of the Schr$\ddot{o}$dinger equation in the be-able dynamics of individual systems in the relevant $\psi$-ensemble. 

The operative equation in be-able dyna\-mics is the guidance equation (30) or, at the level of the stochastic process $\xi$, Eq(44). These equations serve to determine the invariance properties of be-able dynamics; they, however, do not provide clues for possible loss of symmetry in going from observable to be-able dynamics. This role is played by Eq(39) to which one should look for guidance in this matter. 

 We recall from section 3.2 that Bohmian trajectories are de\-termined by both, the evolu\-tion law of the Schr$\ddot{o}$dinger equation and the initial wave function $\psi_{0}$; the latter affects the invariance properties of be-able dyna\-mics through the probability measure $|\psi_0|^{2}$dx appearing in Eq(39). It follows that, given a co\-variance group G of the Schr$\ddot{o}$dinger equation, the corresponding invariance group $G_{b}$ of the be-able dynamics is expected to be the largest subgroup of G leaving the measure  $|\psi_0|^{2}$dx invariant. 
 
 \par
 \noindent \textsl{Remark} While applying this criterion, one should take into consideration the fact that the initial time $t_{0}$ (often taken 0) used in defining $\psi_{0}$ is a fixed parameter in Eq(39). 
 
 For a nonrelativistic spinless free particle, G is the Galilean group. In this case, direct verification of the invariance properties of the guidance equation ([5], section 3.11.1) shows that $G_{b}$ = G. It is worth checking this by the general procedure given above. The transformation under G changes $\psi$ by a multiplicative phase factor (supplementing its transformation as a complex scalar field) ([5], \emph{ibid}). With time fixed, the Galilean boosts act on the Cartesian coordinates $x^{j}$ effectively as space translations. Recalling the remark above, the measure $|\psi_0|^{2}$dx is seen to be G-invariant. 
 
 Such invariance generally does not hold in relativistic situations implying violation of Lorentz invariance in be-able dynamics. At the ensemble level, there is no problem because there one has to deal only with relativistic evolution equations. 
 
 In the paper with the title ``Can Bohmian mechanics be made relativistic ?''[18], the authors show concern/unhappiness at the dBBT being relativistic `only at the relatively superficial level of empirical predictions' and not `in a serious or fundamental sense' and explore possibilities for achieving the supposedly more desirable objective. They end up saying that they are not able to answer the question in the title. We refer to [18] for some related works. 

\par 
\noindent \textsl{Remarks} (i) Principles of relativity were formu\-lated, both in classical and quantum physics, at the observational level. In classical physics, this was so because all physics was essentially at the observational level. Quantum mechanics was developed to provide an explanatory framework for observations relating to processes in the atomic domain. Relstivistic transformation laws were defined for the central objects in this framework --- states and observables. The special feature of Bohmian trajectories for a multiparticle system in which the velocity of an individual particle at any instant depends on the instantaneous positions of all the particles in the system, appears to demand validity of absolute simultaneity in the `Bohmian' domain. According to the present author's understanding, there is no  compelling reason to believe that Lorentz invariance should hold in this domain. It is quite possible that QM is an approximation to a deeper theory (working in terms of be-ables) which has an automorphism group which leads to a relativity group in appropriate approximation/limit. 
 
 Some idea of such a deeper theory may be had from the developments in section 2.1. There we started with an action functional describing be-able dynamics and went on to describe ensemble dynamics in terms of time dependent fields in configuration space --- the kind of objects used also in Schr$\ddot{o}$dinger wave mechanics. The be-ables in the deeper theory envisaged above are expected to be related to some fundamental constituents of matter which need not be particles; they may be fields, strings or more general objects. 

\par 
\noindent (ii) To do proper justice to the word `implementation' in the opening para above, we must consider, in the case when $G_{b}$ is a proper subgroup of G, the implementation of G in be-able dynamics. In a somewhat similar situation in quantum field theory and many body physics, when a symmetry under a group G is spontaneously broken to a proper subgroup K of G (more generally, when only the symmetry under K is unitarily implementable), a standard way of implementing G is in terms of a Hilbert bundle [19,20] in which the fiber is a Hilbert space admitting a unitary action of K and base the space G/K. Different points of the base correspond to different modes of implementation of K. Elements of G outside K relate various modes. 
 
 In the situa\-tion at hand involving the group pair $(G, G_{b})$, it is useful to describe a quantum system in terms of a `quantum triplet' $ ( \mathcal{H}, \mathcal{D}, \mathcal{A}_{q})$ where $\mathcal{H}$ is the system Hilbert space, $\mathcal{D}$ a dense subspace of $\mathcal{H}$ ( supposedly home of Schr$\ddot{o}$dinger wave functions) and $\mathcal{A}_{q}$ an algebra of observables consisting of operators mapping $\mathcal{D}$ into itself. The group G is supposed to act through unitary operators on $\mathcal{D}$ (mapping it onto itself) and $\mathcal{A}_{q}$. 
 
 The type of work in sections 3 and 4 generally involves a complete set $\mathcal{C}$ of observables like the one in Eq(49) and a commutative subalgebra $\mathcal{B}$ of $\mathcal{A}_{q}$ (contained in $\mathcal{C}$) like the one generated by $ (X_1, X_2, X_3, S_3)$ in section 4. The action of $G_{b}$ maps $\mathcal{B}$ onto itself. Action of elements of G outside $G_{b}$ are expected to relate different pairs $(\mathcal{C}, \mathcal{B})$. 
 
 Detailed consideration of these matters in a general setting deserves a separate full length paper; it is, for the time being, left open.
 
 \section{Bohmian trajectories for particle position and spin related to Pauli's equation} 
 
 Pauli's equation is the wave mechanical equation for a nonrelativis\-tic charged spin 1/2 particle interacting with external electromagnetic fields [5] : 
 \begin{equation} 
 i \hbar \frac{\partial \psi}{\partial t} = \left[ - \frac{\hbar^{2}}{2m} ( \mathbf{\nabla} - \frac{ie}{\hbar c} \mathbf{A} )^{2}  + V + \mu \mathbf{B}.\mathbf{\sigma}\right] \psi \equiv H \psi 
 \end{equation} 
 where the usual e$A_{0}$ term has been absorbed in V. In [5], the Bohmian trajectory for the particle position is treated in detail along traditional Bohmian lines. For the treatment of spin related be-ables, however, there is no satisfactory general procedure in dBBT. We shall treat the spin $1/2$ case in the present context following the general procedure outlined in section 3.4. 
 
 Recalling the complete set of Eq(49), we need to consider the commutative algebra generated by the operators $(\{X_{j}; j = 1,2,3\}, S_{3} )$. The spectral spce of these operators is 
 \begin{equation} 
 \Sigma = \mathbb{R}^{3} \times \{ \pm \hbar/2 \} \equiv \mathbb{R}^{3} \times Y_{2};
 \end{equation} 
 a general point of $\Sigma$ is a pair (x,$\lambda$) where $x \in \mathbb{R}^{3}$ and $ \lambda = \pm \hbar/2$. We adopt, on $\Sigma$, the measure $ \mu = \mu_{L} \times \mu_{c}$ where $\mu_{L}$ is the Lebesgue measure on $\mathbb{R}^{3}$ and $\mu_{c}$ is the counting measure on $Y_{2}$. A typical integral on $\Sigma$ will be written as 
 \begin{equation} 
 \int_{\Sigma} f(x, \lambda) d\mu = \sum_{\lambda} \int_{\mathbb{R}^{3}} f(x, \lambda) dx. 
 \end{equation} 
 The two components $\psi_{\alpha}(x,t)(\alpha = 1,2)$ of the wave function in Eq(50) will be written as $\psi(x, \lambda,t)$ with $\lambda = \hbar/2, -\hbar/2$ for $\alpha$ = 1,2 respectively. 
 
 Given a wave function $\psi_{0}(x,\lambda) = \psi(x, \lambda; 0) $ where $\psi(x,\lambda; t)$ is a solution of the Pauli equation (50), the presently relevant probability space analogous to $\mathcal{M}_{0}$ 0f Eq(37) is 
 \begin{equation} 
 \mathcal{M}^{\prime} = ( \Sigma, Borel(\Sigma), |\psi_{0}(x,\lambda)|^{2}d\mu ).  
 \end{equation} 
 Corresponding to the observables $(\{X_{j}\}, S_{3})$ we introduce the stochastic process pair $ \xi(t) = (\xi_{X}(t),\xi_{S}(t))$ where the two processes take values in $\mathbb{R}^{3}$ and $Y_{2}$ respectively. 
 
 For the process $\xi_{X}(t)$, the analogues of equa\-tions (39,40) are 
 \begin{eqnarray} 
 E[\xi_{X}^{j}(t)] & = & \sum_{\lambda} \int\xi_{X}^{j}(t,y,\lambda) |\psi_{0}(y,\lambda)|^{2}dy \nonumber \\
 & = & \langle\psi_{H}|X_{H}^{j}(t)|\psi_{H}\rangle = \langle\psi_{S}(t)|X_{S}^{j}|\psi_{S}(t)\rangle \nonumber \\
 & = & \sum_{\lambda} \int x^{j} |\psi(x, \lambda,t)|^{2}dx 
 \end{eqnarray} 
 implying the probability density 
 \[ \rho_{\xi_X} (x, \lambda,t) = |\psi(x, \lambda,t)|^{2}. \]  
 Note that right hand side of Eq(54) can also be written as $\int x^{j} \tilde{\rho}_{\xi_{X}}(x,t)dx$ where 
 \begin{equation} 
 \tilde{\rho}_{\xi_{X}}(x,t) = \sum_{\lambda}|\psi(x,\lambda,t)|^{2} = (\psi^{\dagger}\psi)(x,t) 
 \end{equation} 
 where, in the last expression, $\psi$ is the usual two-rowed column vector. 
 
 After some calculations similar to (but more lengthy) those leading to equations 42-44, we have 
 \begin{equation} 
 E[\xi_{X}^{j}(t + dt) - \xi_{X}^{j}(t)] = R dt + O(dt^{2}). 
 \end{equation} 
 Here
 \begin{eqnarray*}  
R & = & -  \int x^{j} (\mathbf{\nabla}.\mathbf{K})(x,t)dx   \\
  & = &  \int \left( \frac{K^{j}}{\tilde{\rho}_{\xi_{X}}} \right) (x,t) \tilde{\rho}_{\xi_{X}}(x,t) dx  
 \end{eqnarray*} 
 where 
 \begin{equation} 
 \mathbf{K} = (2im\hbar)^{-1} [\psi^{\dagger}\mathbf{\nabla}\psi -(\mathbf{\nabla} \psi^{\dagger})\psi] - \frac{e}{m\hbar}\mathbf{A} \psi^{\dagger}\psi. 
 \end{equation} 
 Defining the velocity field 
 \begin{equation}
 \mathbf{v}_{X}^{(\psi)}(x,t) = (\mathbf{K}/\tilde{\rho}_{\xi_{X}})(x,t),
 \end{equation}  
  we immediately have, from Eq(56), the ana\-logue of Eq(43), which gives 
 \begin{equation} 
 \frac{d\xi_{X}(t)}{dt} = v_{X}^{(\psi)}(\xi_{X}(t),t). 
 \end{equation} 
 The ve\-locity field in Eq(58) agrees with that of Eq(9.3.8)in [5] to which we refer for an instructive detailed treatment of Bohmian trajectories based on equations (58,59). 
 
 Proceeding as above for the process $\xi_{S}(t)$, we have, using the partition of unity $I = \sum_{\lambda} \int |y,\lambda\rangle dy \langle y, \lambda|$, 
 \begin{eqnarray} 
 E[\xi_{S}(t)] & = & \sum_{\lambda} \int \xi_{S}(t; y,\lambda) |\psi_{0}(y, \lambda)|^{2} dy  \nonumber \\
               & = & \langle\psi_{H}| S_{3}^{H}(t) |\psi_{H}\rangle \nonumber \\
               & = & \langle\psi_{S}(t)|S_{3}|\psi_{S}(t)\rangle   \nonumber \\
               & = & \sum_{\lambda} \int dy \lambda |\psi(y,\lambda,t)|^{2} \nonumber \\
               & \equiv & \sum_{\lambda} p_{\lambda}^{(\psi)} 
 \end{eqnarray} 
 where $p_{\lambda}^{(\psi)}$ is, given the state $\psi$, the probability of the observable $S_{3}$ \emph{having} the value $\lambda$. [Caution : The subscript S in $\xi$ and $\psi$ above has different meanings.] Now 
 \begin{equation} 
 E[\xi_{S}(t+dt) - \xi_{S}(t)] = Q dt + O(dt^{2}) 
 \end{equation}
 where 
 \begin{eqnarray*}
 Q  & = & (i\hbar)^{-1}\langle\psi_{S}(t)|[S_{3}, H]|\psi_{S}(t)\rangle  \\
    & = &  \sum_{\lambda}\int dy J_{S}^{(\psi)}(y, \lambda; t).   
 \end{eqnarray*} 
 Here 
 \[J_{S}^{(\psi)}(y,\lambda; t) = - \frac{i\mu}{2}\psi^{*}(y,\lambda; t)(B_{-}\sigma_{+} + B_{+}\sigma_{-}) \psi(y,\lambda; t) \]
 and we have used the relation 
 \begin{displaymath}
 [ S_{3}, H] = \mu [ S_{3}, \mathbf{B}.\mathbf{\sigma}] = \frac{\mu \hbar}{2} (B_{-} \sigma_{+} + B_{+} \sigma_{-}) 
 \end{displaymath} 
 with $B_{\pm} = (B_{1} \pm B_{2}), \sigma_{\pm} = (\sigma_{1} \pm  \sigma_{2})/2.$ Defining 
 \[ v_{S}^{(\psi)}(y,\lambda; t) = J_{S}^{(\psi)}(y,\lambda; t)/|\psi(y,\lambda; t)|^{2}, \] 
 we have 
 \begin{equation} 
 E[ \xi_{S}(t+dt) - \xi_{S}(t) - v_{S}^{(\psi)}(\xi_{S}(t),t)dt] = O(dt^{2})      
\end{equation} 
showing differentiability of $\xi_{S}$ with 
\begin{equation} 
\frac{d\xi_{S}(t)}{dt} = v_{S}^{(\psi)}( \xi_{S}(t),t); 
\end{equation} 
the corresponding trajectory function satisfies the guidance relation 
\begin{equation} 
\frac{\partial \xi_{S}(t; y,\lambda)}{\partial t} = v_{S}^{(\psi)}(\xi_{S}(t; y,\lambda),t). 
\end{equation} 

In the Stern-Gerlach (S-G) experiment, one has $\mathbf{B} = B \mathbf{e}_{3}$ along z-direc\-tion, implying $v_{S}^{(\psi)}$ = 0. It follows that, a Bohmian trajectory starting at time t=0 at the point (y,$\lambda$) in the space $\Sigma$, will continue to have the fixed value $\lambda$ for $S_{3}$ for all t. The experimental outcome (in view of lack of control on initial configurations of atomic systems) will, therefore, reflect the distribution of $\lambda$ as given by the initial wave function $\psi_{0}$, in complete agreement with the Born rule.  

\par
\noindent \textsl{Remarks} (i) The treatment above provides the framework for be-able based analysis of the case when the magnetic field in the S-G experiment is taken in a general direction. 

\par
(ii) The treatment of spin presented above followed the general procedure outlined in section 3.4 and can be easily extended to other spin values.  

\par
(iii) In the book [5], the author has employed the `spin vector field'  
\begin{equation}                
\mathbf{s}(x,t) = \left( \psi^{\dagger} (\frac{\hbar}{2} \mathbf{\sigma} \psi )/ (\psi^{\dagger}\psi) \right)(x,t) 
\end{equation} 
which is the `local expectation value' [recall Eq(48)] for the op\-erator $ \mathbf{S} = \frac{\hbar}{2} \mathbf{\sigma}$. The value of this field on a Bohmian trajectory, called `spin vector' of the particle, has been used in [5] to provide the traditional Bohmian argument for the outcome in  the S-G experiment. This involves instructive physics; the overall reasoning, however, has the usual `hidden variable' aspect of dBBT and does not integrate well with traditional Hilbert space based QM.
 
\section{Derivation of von Neumann projec\-tion rule based on S-B evolution of individual systems} 

As promised in section 3.1, we shall now derive von Neumann's projection equation (29) employing Schr$\ddot{o}$dinger-Bohm evolu\-tion of individual systems in a measurement situation. 

In the treatment of measurements in section 2.2, a state vector $|\Psi\rangle$ in the system-apparatus Hilbert space $ \mathcal{H} = \mathcal{H}_{\mathcal{S}} \otimes \mathcal{H}_{\mathcal{A}}$ represents an ensemble of (object system)-apparatus systems. Considering individual systems in this en\-semble along the lines of section 3.1, we have stochastic processes $\xi_{X}(t)$ and $\eta_{Y}(t)$ (where X and Y are configuration observables in the configuraration spaces M and $M^{\prime}$ of the system $\mathcal{S}$ and apparatus$\mathcal{A}$) in the probability space [recall Eq(37) above] 
\begin{equation} 
\mathcal{M} = \left( M \times M^{\prime}, Borel(M \times M^{\prime}), \mu_{M} \times \mu_{M^{\prime}} \right). 
\end{equation} 
\noindent \textsl{Note} In the derivation given below, no serious use is made of the manifold natures of M and $M^{\prime}$ and the measures $\mu_{M}$ and $\mu_{M^{\prime}}$; for simplicity, one may take M = $\mathbb{R}^{N}$ and $M^{\prime} = \mathbb{R}^{N^{\prime}}$ with respective Lebesgue measures, \newline 
\noindent Sample paths $\xi_{X}(t,x_{0})$ and $\eta_{Y}(t,y_{0})$ of these processes corresponding to points $x_{0} \in M$ and $y_{0} \in M^{\prime}$ satisfying the conditions [see Eq(38)] 
\begin{equation} 
\xi_{X}(0,x_{0}) = x_{0}, \ \ \eta_{Y}(0,y_{0}) = y_{0} 
\end{equation}      
represent Bohmian trajectories of system and apparatus starting at points $x_{0}$ and $y_{0}$ respectively; we shall refer to it below as ($x_{0}, y_{0}$) trajectory. 

The initial state $\Psi_{in}$ of Eq(25)represents an ensemble of (system + apparatus) systems. For the individual system in the $\Psi_{in}$-ensemble corresponding to the $(x_{0}, y_{0})$ trajectory, the state at t = 0 is 
\begin{equation} 
(|\Psi_{in}\rangle; x_{0}, y_{0}) = \left(|\Psi_{in}\rangle; \xi_{X}(0, x_{0}), \eta_{Y}(0,y_{0}) \right). 
\end{equation} 
Corresponding to the ensemble evolution 
\[ U(t) |\Psi_{in}\rangle = |\Psi(t)\rangle \] 
we have the evolution of the state (68) represented by the obvious extension V(t) of U(t) given by 
\begin{equation} 
V(t) (|\Psi_{in}\rangle; x_{0}, y_{0} ) = \left( |\Psi(t)\rangle; \xi_{X}(t, x_{0}), \eta_{Y}(t, y_{0}) \right). 
\end{equation} 

Recalling the macroscopic distinguishability of pointer positions corresponding to various j in Eq(26) [implying the almost non-overlap of the corresponding wave packets $\phi_{j}(y)$], location of the $(x_{0}, y_{0})$ trajectory at the final time t = T 
\[ \left( \xi_{X}(T,x_{0}), \eta_{Y}(T,y_{0}) \right) \in M \times M^{\prime} \] 
chooses precisely one of the terms in Eq(26) thereby ensuring a unique outcome in a single run of the experiment. 

Generally more than one $y_{0} \in M^{\prime}$ may correspond to the same j. Moreover, a choice of an initial point $y_{0} \in M^{\prime} $ leads to a unique Bohmian trajectory for the apparatus. This implies a partition of $M^{\prime}$ into disjoint domains $D_{j}^{\mathcal{A}}$ corresponding to different j : 
\begin{equation} 
D_{j}^{\mathcal{A}} = \{ y_{0} \in M^{\prime}; \eta_{Y}(T, y_{0}) \in supp (\phi_{j}) \}. 
\end{equation} 
Noting from Eq(26) that, corresponding to a given j, we have some Bohmian system $\mathcal{S}$ trajectories, we have similar decomposition of M into domains 
\begin{equation} 
D_{j}^{\mathcal{S}} = \{ x_{0} \in M; \xi_{X}(T, x_{0}) \in supp(\psi_{j}) \}. 
\end{equation} 

Let $ K_{j} = D_{j}^{\mathcal{S}} \times D_{j}^{\mathcal{A}} $. Using these domains, we express the initial (system + apparatus) density operator as a sum of `partial density operators'  : 
\begin{equation} 
\rho^{(\Psi_{in})} = \sum_{j} \rho^{(\Psi_{in})}_{j} 
\end{equation} 
where 
\begin{equation} 
\rho^{(\Psi_{in})}_{j} = E \left[ \chi^{(K_{j})} \left( \xi_{X}(0,x_{0}), \eta_{Y}(0, y_{0}) \right) \right] \rho^{(\Psi_{in})}. 
\end{equation} 
Here $\chi^{(K_{j})}$ is the characteristic function of the domain $K_{j} \subset \-(M \times M^{\prime})$ and, recalling Eq(39), the expectation E is defined as [noting that $\xi_{X}(0,x_{0}) = x_0, \eta_{Y}(0,y_{0}) = y_0 $ ] 
\begin{equation} 
E[f(x_0,y_0)]  = \int_{M\times M^{\prime}} f(x,y) |\Psi_{in}(x,y)|^{2} d\mu_{M}(x) d\mu_{M^{\prime}}(y). 
\end{equation} 
Note that 
\begin{eqnarray} 
\sum_{j} E[\chi^{(K_{j})}(x_0,y_0)] & = &   \Sigma_{M \times M^{\prime}} |\Psi_{in}(x_{0}, y_{0})|^{2}   \nonumber  \\
                                    & = & 1
\end{eqnarray} 
where 

\[ \Sigma_{M \times M^{\prime}} (.) = \int_{M \times M^{\prime}} (.) d\mu_{M}(x_{0}) d\mu_{M^{\prime}}(y_{0}). \]

Eq(75) plus the fact that the S-B evolution does not distinguish points in a single domain $K_j$, justifies equations (72,73).

The evolution map V(t) of Eq(69) induces the evolution map W(t) of a partial density operator such that
\begin{equation} 
W(t) \left( \rho^{(\Psi_{in})}_{j} \right) = A(t) U(t) \rho^{(\Psi_{in})} U(t)^{\dagger} 
\end{equation} 
where

\[ A(t) = E\left[\chi^{(K_{j})} \left(\xi_{X}(t,x_{0}), \eta_{Y}(t,y_{0}) \right) \right]. \]

Putting t = T in this equation, we have, recalling Eq(26), 
\begin{equation}
W(T) [\rho^{(\Psi_{in})}_{j}] = E\left[\chi^{(K_{j})}\left(\xi_{X}(T,x_{0}), \eta_{Y}(T,y_{0}) \right) \right] \rho^{(\Psi_{f})}. 
\end{equation} 
Now 
\begin{equation}
\chi^{(K_{j})}\left( \xi_{X}(T,x_{0}), \eta_{Y}(T,y_{0}) \right) = \left[ \begin{array}{ll} 
1 & \mbox{ if condition B holds} \\
0 & \mbox{otherwise.} \end{array} \right.
\end{equation}
Here condition B means 
\[ \xi_{X}(T,x_{0}) \in supp(\psi_{j}), \eta_{Y}(T,y_{0}) \in supp(\phi_{j}). \]
Recalling Eq(27) for $\Psi_{f}(x,y)$ and the probability interpretation for the wave function, we have 
\begin{equation} 
E\left[ \chi^{(K_{j})} \left( \xi_{X}(T,x_{0}), \eta_{Y}(T,y_{0}) \right) \right] = \left[ \begin{array}{ll} 
1 & \mbox{if condn C holds}  \\ 
0 & \mbox{otherwise.} \end{array} \right.
\end{equation}   
Here condition C means that at $t=T$, the X and Y configu\-rations are, respectively, in supp($\psi_{j}$) and supp ($\phi_{j}$).   
Equations (77,79) now give, recalling the notation in Eq(29), 
\begin{equation} 
W(T) \left( \rho^{(\Psi_{in})}_{j} \right) = P_{j} \rho^{(\Psi_{f})} P_{j}. 
\end{equation} 
Finally, equations (72,80) give 
\begin{equation} 
W(T) (\rho^{(\Psi_{in})}) = \sum_{j} P_{j} \rho^{(\Psi_{f})} P_{j}. 
\end{equation} 
Note that right hand side of Eq(81) is precisely the density operator $\rho_{red}$ of Eq(29). This completes the promised derivation of the von Neumann projection rule.  

\par
\noindent \textsl{Remarks} (i) In Eq(81), the operator W(T) represents the overall effect, at the ensemble level, of the Schr$\ddot{o}$dinger-Bohm evolutions of individual (system + apparatus) systems along  their Bohmian trajectories. On the other hand, the operator U = U(T) of Eq(26) represents the traditional Schr$\ddot{o}$dinger evolution [recalling Eq(28)] 
\[ U(T) \rho^{(\Psi_{in})} U(T)^{\dagger} = \rho^{(\Psi_{f})}. \] 
It is the difference between the actions of W(T) and U(T) that corresponds to the `discontinuous and non-causal process' envisaged in von Neumann's projection postulate [1]. 

What is the physics behind this difference ? Putting the question differently, what is missing in the action of U(T) at the ensemble level ? Key to the answer lies in macroscopic distinguishability of pointer positions in the apparatus [21] which leads to the break up of the space $M^{\prime}$ into domains $D_{j}^{\mathcal{A}}$ of Eq(70). The Schr$\ddot{o}$dinger-Bohm evolution of individual systems which is reflected in the action of W(T) in Eq(80), distinguishes between these domains; the operator U(T) obviously does not. 

\par
\noindent (ii)  Environment induced decoherence is, under plausible assump\-tions, known to lead to the reduced density operator of Eq(29) [22, 23, 15]. To see its relevance in the present context, note that, in the derivation above, the whole configuration spaces M and $M^{\prime}$ of the system($\mathcal{S}$) and apparatus ($\mathcal{A}$) are involved. The degrees of freedom of the system and the apparatus which do not participate in the measurement interaction constitute their respective `internal environments'[15]. It is the decoherence effect of the internal environments which is included in the action of W(T) but not of U(T); vehicles of this decoherence effect are the relevant Bohmian trajectories. Since the measured system is generally microscopic, its internal environment is relatively insignificant; the main contribution comes from the internal environment of the apparatus. Moreover, since, ideally, only the whole universe is a closed system, replacing, in the treatment of measurements above, `apparatus' by `rest of the universe', one has the external environment also included (essentially without any extra cost). 

To see the environment induced decoherence in `live action' in the present context in the setting of noncommutative geometry, see [21]. 

\par
\noindent (iii) It needs to be emphasised that a superposition of `macro states' like that in Eq(26) \emph{never} appears in a proper description of concrete physical systems involved in single runs of experiments. These systems follow appropriate Bohmian trajectories which, as shown above, always pick up a single term in the expansion (26). [ To grasp this point better, it is helpful to recall the parallel situation in H-J theory treated in section 2.1. Recalling Eq(5) and the development around it, it is clear that a surface like $S(q, t_{0})$ = constant never appears in the description of evolution of single systems which follow trajectories described by equations 5-7.] 

A need of consideration of `inactive packets' [ the terms in Eq(27) corresponding to j values not picked up by (system + apparatus) Bohmian trajectories in individual runs of experiments] arises in traditional treatments of Bohmian mechanics (see, for example, [4] section 6.1) because, in these treatments, the Schr$\ddot{o}$dinger wave function is assumed to be a physical field analogous to the electromagnetic field. We have given arguments in section 2.3 as to why this assumption is not acceptable. 

\section{Concluding Remarks} 

(1) \textsl{Summing up :} Working consistently in a Hilbert space based probabilistic framework and adopting ensemble interpretation of the Schr$\ddot{o}$dinger wave function, we have augmented the traditional state-observable framework (which serves for ensemble stochastics) by making use of the probability measure defined on the system configuration space by a fixed time wave function $\psi$ for the treatment of stochastics of individual systems in the $\psi$-ensemble; this is done by employing stochastic processes in this probability space corresponding to various Heisenberg operators in the system Hilbert space. Sample paths of the stochastic process corresponding to the Heisenberg position operator represent configuration space trajectories of individual systems in the $\psi$-ensemble. Consistency between the individual and ensemble statistics (enforced in terms of equality of appropriate mean values) implies that these trajectories are Bohmian trajectories. Bohmian trajectories corresponding to spin observables are treated by extending the configuration space (which is the spectral space of position operators) to include the spectral space of appropriate spin operators. A straightforward derivation of von Neumann projection formula for the (system + apparatus) density operator based on Schr$\ddot{o}$dinger- Bohm evolution of individual systems along their Bohmian trajectories is given. The resulting augmented framework has (besides retaining all the positive features of the traditional Hilbert space based formalism) all the positive features of dBBT and none of its negative ones. 

A positive implication of the developments described above is the demonstration of appropriateness of the ensemble interpretation of  Schr$\ddot{o}$dinger wave function. 

\par
\noindent (2) In the traditional development of dBBT, primary role of the wave function $\psi$ is assumed to be that of guidance through a guidance equation like (30); its role in providing the probability measure $|\psi(x)|^{2}dx$ (referred to in some works as `quantum equilibrium distribution' [24,17,16]) is considered as secondary. Some authors have considered the possible role of `quantum non-equilibrium'(represented by a probability measure $\rho(x)dx$ eventually converging to $|\psi(x)|^{2}dx$) prevailing in situations like dynamics of matter in the early universe. (For details and references see the review [25].) In the light of the present work, the proper conclusion appears to be that, so long as an appropriate Schr$\ddot{o}$dinger equation is taken to be operative, observational evidence is unlikely to support the conclusions based on consideration of quantum non-equilibrium. 

\par
\noindent (3) Some autmhors, specially Bell ([26],p. 117) consider the place of measurements as relatively insignificant at the fundamental level : 
\begin{quote}
`` The concept of `measurement' becomes so fuzzy on reflection that it is quite surprising to have it appearing in physical theory \emph{at the most fun\-damental level''.} 
\end{quote}
The fuzziness mentioned in the quote above has been, hopefully, substantially reduced by the work presented in sections 3,5 above. Measurements are controlled observations involving apparatus systems having macroscopically distinguishable pointer subsystems. In fact, no physics and, therefore, no science is possible without such systems because human brain acts as such an apparatus in all scientific activity. Measurement like processes are all the time going on (in the multitude of human brains) not only in scientific but all human activity.    

\par
\noindent (4) The augmented Hilbert space based formalism presented above has potential for application in all theoretical work employing Schr$\ddot{o}$dinger equation with traditional probability interpretation for the wave function. 

\par
\noindent (5) With the provision for the treatment of individual systems now available in the Hilbert space based formalism augmented as above, the ensemble interpretation of Schr$\ddot{o}$dinger wave function is no longer an obstacle for its application in the  treatment of quantum mechanics of the universe. A special attraction is the availability of objec\-tively real trajectories in a consistent probabilistic framework. Recalling the discussion in section 3.2, a Bohmian trajectory of the universe is expected to be determined by the action function for the matter-gravity system and the initial wave function of the universe (and, of course, its initial configuration in the configuration space of the universe.) We have, therefore, a framework for studying quantum evolution as well as initial conditions relating to the universe. Moreover, since it avoids the extravagance of introducing a multi\-tude of universes, it has some advantage over the so-called `many worlds'formalism [27].

In this context, the problem of time in quantum gravity acquires special significance due to absence of $\partial \Psi / \partial t$ term in the Wheeler-DeWitt equation [28], the prevalent `evolution equation' for the universal wave function $\psi$. In the review [29] devoted to application of dBBT in quantum cosmology, this problem is taken care of by introducing an internal/relational time T as a function of some field variables to obtain a Schr$\ddot{o}$dinger like equation for $\Psi$ [Equations (23, 25) in [29]]. An impressive result in that work is, in the perfect fluid model for matter in the universe, Eq (60) for the cosmological scale function a(T) which is a bounce solution without the big bang singularity.     

In some works, one tries to intuitively identify appropriate velocity fields [generally from the equation obtained by writing $\Psi = |\Psi|exp(iS/\hbar)$ in the Wheeler-DeWitt equation and taking the imaginary part] and write the guidance equations for the relevant be-ables `by hand'. [See, for example, equations (88,89) in [25].] In the light of the present work, internal consistency of such a procedure appears to be suspect. 
    
\vspace{3mm} 
\noindent \textbf{Address} Flat No. 704, Block 7, Eastend Apartments, Mayur Vihar Phase I Extension, Delhi 110096, India 

\noindent \textbf{Email} tulsi@iitk.ac.in 

\vspace{2mm} 
\noindent \textbf{References} 
\begin{description}
\item[[1]] J.von Neumann : Mathematical Foundations of Quan\-tum Mechanics, Princeton University Press (1955) 
\item[[2]] L. de Broglie : The new dynamics of quanta, 1927 Solvay Conference, arXiv : quant-ph/0609184
\item[[3]] D. Bohm : A suggested interpretation of the quantum theory in terms of ``hidden'' variables I,II , Physical Review \textbf{85}, 155-179, 180-193 (1952) 
\item[[4]] D. Bohm, B.J. Hiley : The Undivided Universe. \emph{An Ontological Interpretation of Quantum Theory}, Routledge, London (1995) 
\item[[5]] P.R. Holland : The Quantum Theory of Motion, Cambridge University Press, Cambridge (1995) 
\item[[6]] J.S. Bell : Beables for quantum field theory, CERN-TH 4035/84; Chapter 19 in his book `Speakable and Unspeak\-able in Quantum Mechanics', Cambridge University Press (1987) 
\item[[7]] R.P. Feynman : Spacetime approach to non-relativis\-tic quantum mechanics, Reviews of Modern Physics \textbf{20} 367-387 (1948) 
\item[[8]] Tulsi Dass : Dynamical probability, particle trajectories and completion of traditional quantum mechanics, arXiv : quant-ph/0505190 
\item[[9]] F. Guerra : Structural aspects of stochastic mechanics and stochastic field theory, Physics Reports \textbf{77} 263-312 (1981) 
\item[[10]] Tulsi Dass : Symmetries, Gauge Fields, Strings and Fundamental Interactions,volume I : 
Mathematical Techniques in Gauge and String Theories, Wiley Eastern Limited (1993) 
\item[[11]] A. Papoulis : Probability, Random Variables and Stochas\-tic Processes, Mc Graw-Hill Internatio\-nal Book Company (1965) 
\item[[12]] Tulsi Dass : Towards an autonomous formalism for quantum mechanics, arXiv : quant-ph/0207104 
\item[[13]] J.A. Wheeler, W.H. Zurek : Quantum Theory and Measurements, Princeton University Press (1983) 
\item[[14]] M. Jammer : The Philosophy of Quantum Mechanics, Wiley, New York (1974) 
\item[[15]] Tulsi Dass: Measurements and Decoherence, arXiv: quant-ph/0505070 
\item[[16]] D. D$\ddot{u}$rr, S. Teufel : Bohmian Mechanics.\emph{The Phy\-sics and Mathematics of Quantum Theory}, Springer-Verlag, Berlin (2009) 
\item[[17]] D. D$\ddot{u}$rr, S. Goldstein, N. Zanghi : Quantum equilibrium and the role of operators as observables in quantum theory, J. Stat. Phys. \textbf{116} 959-1055 (2004) 
\item[[18]] D. D$\ddot{u}$rr, S. Goldstein, T. Norsen, W. Struyve, N. Zanghi : Can Bohmian Mechanics be made relativistic ?, Proceedings of the Royal Society A \textbf{470} 20130699 (2014) 
\item[[19]] H.J. Borchers, R.N. Sen : Relativity groups in the presence of matter, Comm. Math. Phys. \textbf{42} 101-126 (1975) 
\item[[20]] S.K. Bose : Broken symmetry and bundle representations, Lett. al Nuovo Cimento \textbf{28} 146-150 (1980)
\item[[21]] Tulsi Dass : A stepwise planned approach to the solution of Hilbert's sixth problem. III : Measurements and von Neumann's projec\-tion/coll\-apse rule, Pramana \textbf{77} 1031-1051 (2011) 
\item[[22]] W.H. Zurek : Decoherence, einselection and quantum origin of the classical, Rev. Mod. Phys. \textbf{75} 715 (2003); arXiv : quant-ph/0105127 
\item[[23]] D. Giulini et al (ed): Decoherence and the Appearance of a Classical World in Quantum Theory, Springer (1996); second edition [Joos et al (ed), (2003)] 
\item[[24]] D. D$\ddot{u}$rr, S. Goldstein, N. Zanghi : Quantum equilibrium and the origin of absolute uncertainty, J. Stat. Phys. \textbf{67} 843-907 (1992) 
\item[[25]] A Valentini : De Broglie-Bohm quantum mechanics, Ency\-clopedia of Mathematical Physics, second edition, Elsevier, Amsterdam (2024); arXiv : 2409.\-01294 (quant-ph) 
\item[[26]] J.S. Bell : Quantum mechanics for cosmologists, chapter 15 in the book in Ref [6] 
\item[[27]] B.S. DeWitt, N. Graham (ed) : The Many-Worlds In\-terpretation of Quantum Mechanics, Princeton Uni\-versity Press, Princeton (1973) 
 \item[[28]] C. Kiefer : Quantum Gravity, Oxford University Press, Oxford (2012) 
 \item[[29]] N. Pinto-Neto : The de Broglie-Bohm quantum theory and its application to quantum cosmology, Universe \textbf{7}, 134 (2021); arXiv : 2111.03057 (gr-qc) 
     
\end{description} 
\end{document}